\documentclass[journal=jacsat,manuscript=article]{achemso}

\usepackage[version=3]{mhchem} 
\usepackage{amssymb}

\newcommand\device[1]{$\mathsf{ibm}\_\mathsf{#1}$}
\newcommand\crt[1]{\hat{a}^\dagger_{#1}}
\newcommand\dst[1]{\hat{a}^{\phantom{\dagger}}_{#1}}
\newcommand\bts[1]{{\bf{#1}}}

\usepackage{xcolor}
\newcommand{\nocontentsline}[3]{}
\newcommand{\tocless}[2]{\bgroup\let\addcontentsline=\nocontentsline#1{#2}\egroup}

\title{Implicit solvent sample-based quantum diagonalization}

\author{Danil Kaliakin}
\affiliation{Center for Computational Life Sciences, Lerner Research Institute, The Cleveland Clinic, Cleveland, Ohio 44106, United States}

\author{Akhil Shajan}
\affiliation{Center for Computational Life Sciences, Lerner Research Institute, The Cleveland Clinic, Cleveland, Ohio 44106, United States}
\alsoaffiliation{Department of Chemistry, Michigan State University, East Lansing, Michigan 48824, United States}

\author{Fangchun Liang}
\affiliation{Center for Computational Life Sciences, Lerner Research Institute, The Cleveland Clinic, Cleveland, Ohio 44106, United States}

\author{Kenneth M. Merz Jr.}
\email{kmerz1@gmail.com}
\affiliation{Center for Computational Life Sciences, Lerner Research Institute, The Cleveland Clinic, Cleveland, Ohio 44106, United States}
\alsoaffiliation{Department of Chemistry, Michigan State University, East Lansing, Michigan 48824, United States}

\begin{document}

\begin{tocentry}

    \centering
    \includegraphics[width=\textwidth]{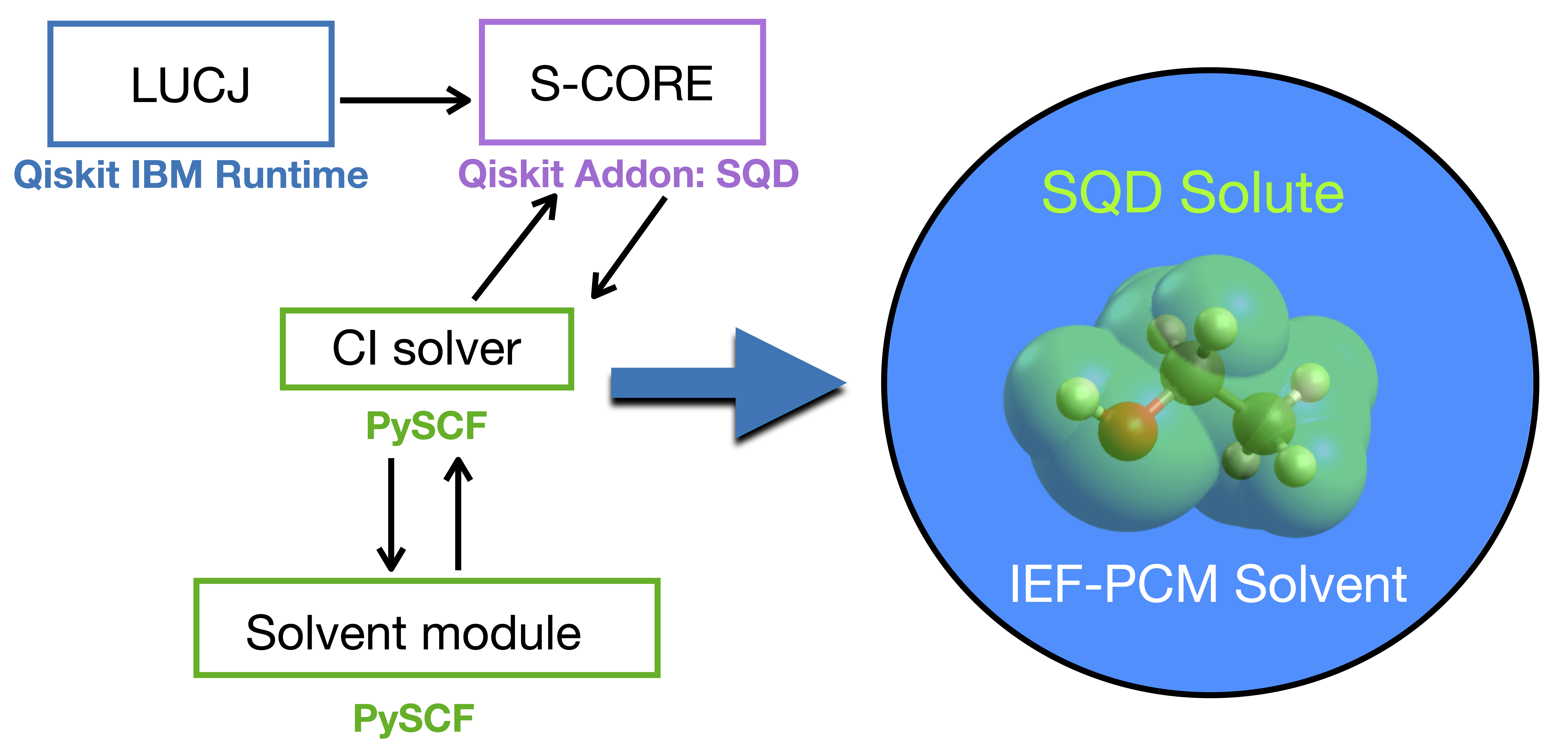}

\end{tocentry}


\begin{abstract}
The sample-based quantum diagonalization (SQD) method shows great promise in quantum-centric simulations of ground state energies in molecular systems. Inclusion of solute-solvent interactions in simulations of electronic structure is critical for biochemical and medical applications. However, all of the previous applications of the SQD method were shown for gas-phase simulations of the electronic structure. The present work aims to bridge this gap by introducing the integral equation formalism polarizable continuum model (IEF-PCM) of solvent into the SQD calculations. We perform SQD/cc-pVDZ IEF-PCM simulations of methanol, methylamine, ethanol, and water in aqueous solution using quantum hardware and compare our results to CASCI/cc-pVDZ IEF-PCM simulations. Our simulations on \device{cleveland}, \device{kyiv}, and \device{marrakesh} quantum devices are performed with 27, 30, 41, and 52 qubits demonstrating the scalability of SQD IEF-PCM simulations. 
\end{abstract}

\maketitle

\tocless\section{Introduction}
Solvation effects are pivotal for a wide range of applications, including drug design\cite{chambers1999modeling, wong2011accounting, yoshida2017role}, protein design\cite{gonzalez2011importance, england2011role, davis2018does, makarov2002solvation} and catalysis\cite{gauthier2017solvation, gounder2013catalytic, gounder2012solvation}, as they influence reaction mechanisms and molecular properties\cite{ben2005solvation, ratkova2015solvation, chang2006recent}.
However, the accurate modeling of solvated chemical systems remains one of the most critical challenges in computational chemistry.
These effects arise from complex solute-solvent interactions, encompassing electrostatics, dispersion, hydrogen bonding, and polarization, which makes the problem inherently many-body in nature\cite{cramer2008universal, cramer1999implicit, tomasi1999ief, tomasi2005quantum, mennucci1997evaluation, mennucci2008continuum}.

Solvation is traditionally addressed using explicit or implicit models\cite{zhang2017comparison, reddy2006implicit}. 
Explicit models simulate individual solvent molecules, capturing detailed solute-solvent interactions, but they require extensive sampling due to the many degrees of freedom involved\cite{kelly2006adding, zhang2001solvent, shen2002long}. 
Implicit models, such as the polarizable continuum model (PCM) and its advanced formulations like IEF-PCM, approximate the solvent as a continuous dielectric medium, reducing computational cost while capturing dominant electrostatic interactions\cite{mennucci1997evaluation, mennucci2008continuum, cances1998new, tomasi1999ief, tomasi2005quantum, herbert2021dielectric, cramer1999implicit, cramer2008universal}.
Despite these advances, integrating implicit solvation models with high-accuracy quantum chemistry methods, such as coupled cluster (CC) theory \cite{caricato2019coupled, caricato2010coupled, cammi2010coupled} and complete active space configuration interaction (CASCI)\cite{seritan2021terachem, boguslawski2011construction, levine2021cas}, which provide systematically improvable treatments of electronic correlation, remains computationally demanding for systems containing tens to hundreds of atoms. The computational costs of these methods scale steeply with system size.

Quantum computing offers a transformative approach to overcome these limitations. 
Unlike classical systems, which encode information as bits, quantum computers leverage qubits that can exist in superpositions of states, enabling efficient representation and manipulation of complex quantum systems. 
Quantum algorithms, including the variational quantum eigensolver (VQE)\cite{fedorov2022vqe, liu2022layer, tang2021qubit, tilly2022variational}, quantum phase estimation (QPE)\cite{parker2020quantum, veis2016quantum, o2019quantum}, and sample-based quantum diagonalization (SQD)\cite{kanno2023quantum, nakagawa2023adapt, robledo2024chemistry,barison2024ext-sqd}, have been developed to solve the electronic Schr\"odinger equation. These methods promise to achieve chemical accuracy for increasingly complex systems as quantum hardware matures.

Recent research has demonstrated the feasibility of integrating quantum computing with different solvent models. 
For example, the use of VQE combined with IEF-PCM \cite{castaldo2022quantum} and the polarizable embedded framework \cite{kjellgren2024variational} has yielded promising results (on classical simulators of quantum circuits) in calculating total energies for small molecules in solution, achieving accuracies comparable to high-level classical methods. 
However, despite these advances in solvation chemistry, VQE and quantum phase estimation-based algorithms face challenges due to their reliance on iterative evaluations of the energy expectation value, which are highly susceptible to noise and measurement errors.
In comparison, SQD bypasses the need for variational optimization by leveraging quantum sampling to construct a reduced Hamiltonian in a subspace, which is then diagonalized classically\cite{robledo2024chemistry}. This approach has demonstrated its effectiveness for covalent molecules, metal-sulfur clusters\cite{robledo2024chemistry}, supramolecular interactions\cite{kaliakin2024accurate}, triplet states\cite{liepuoniute2024quantum}, and excited-state systems\cite{barison2024ext-sqd}. Compared to VQE, SQD is more robust to noise, reduces measurement costs, and is well-suited for near-term quantum devices, making it a promising tool for quantum chemistry applications.

In this study, we integrate the SQD method with implicit solvation models to advance first-principles calculations of solvated systems. Section 2 describes the methodology, detailing the integration of SQD with IEF-PCM, while Section 3 outlines computational details and SQD IEF-PCM code implementation. Section 4 presents results for four polar molecules (water, methanol, ethanol, and methylamine) in aqueous solution, demonstrating that SQD IEF-PCM achieves accuracy comparable to classical high-level methods such as CASCI IEF-PCM, with improved energy convergence as sample sizes increase. The paper concludes with a discussion of the potential of hybrid quantum-classical workflows for practical chemistry applications.


\tocless\section{Methods}

\tocless\subsection{Implicit solvation model}

Below we briefly summarize the key details of the implicit solvation method employed in a present paper. For a detailed description of implicit solvation we refer readers to the review papers on the subject~\cite{mennucci1997evaluation, cances1998new, tomasi1999ief, herbert2021dielectric}. Implicit solvation methods are based on the idea that the interaction of a solute with the solvent can be approximated in a manner where the target subsystem (the solute) is described explicitly including the electronic structure and the molecular geometry, while the secondary subsystem (the solvent) is modeled as an infinite macroscopic continuum medium~\cite{mennucci1997evaluation}. In the resulting model the solute is embedded in a molecular shaped cavity that forms a surface interacting with the surrounding solvent described as a structureless polarizable dielectric. The implicit solvation term is introduced in the Hamiltonian, and in the corresponding Schrödinger equation, through the solute-solvent interaction potential. The corresponding expression can be written as
\begin{equation}
\label{eq:solventH}
[\hat{H}^{0}+\hat{V}_{int}]\Psi=E\Psi,
\end{equation}
here $\hat{H}^{0}$ is the Hamiltonian of the solute in vacuo, $\Psi$ is the solute wavefunction, $\hat{V}_{int}$ is the solute-solvent interaction potential, and $E$ is the total energy of the solute. The formulation of $\hat{V}_{int}$ implies the thermally averaged distribution function of the solvent molecules, formally equating the $E$ to the free energy, $G$, of the given molecule in solution, while the $\hat{V}_{int}\Psi$ equates to the solvation free energy, $G_{solv}$. The $\hat{V}_{int}$ explicitly depends on the solute electronic wave function as
\begin{equation}
\label{eq:interpot}
\hat{V}_{int}(\Psi)=\hat{A}(\Psi\Psi^{*}),
\end{equation}
where $\hat{A}$ is the integral operator. The variational solution of Eq.~\eqref{eq:solventH} requires minimization of $G$ which can be expressed as
\begin{equation}
\label{eq:Gmin}
G(\Psi)=\langle\Psi|\hat{H}^{0}+\hat{V}_{int}'+\frac{1}{2}\hat{V}_{int}''(\Psi)|\Psi\rangle,
\end{equation}
here $\hat{V}_{int}'$ and $\hat{V}_{int}''$ solute-solvent interaction potential corresponding to the current step and its change corresponding to the next iteration. The Eq.~\eqref{eq:Gmin} summarizes the essence of a self-consistent reaction-field (SCRF) problem. The solute's charge distribution both polarizes, and is polarized by, its environment. Which means Eq.~\eqref{eq:Gmin} needs to be iterated to reach self-consistency of both effects~\cite{mennucci1997evaluation, herbert2021dielectric}.

The polarizable continuum model (PCM) is a class of continuum solvation models in which the three-dimensional differential equations of continuum electrostatics are replaced with a two-dimensional boundary-element problem defined on the cavity surface $\Gamma$, where $\Gamma$ represents the interface between atomistic solute and continuum solvent. The integral equation formalism (IEF) of PCM further improves the implicit solvation model by expressing the surface charge, $\sigma(s)$, through the electrostatic potential only, without the need to calculate the derivatives of the electrostatic potential. The utilization of electrostatic potential derivatives can lead to higher sensitivity to discretization errors, which is undesirable. In IEF-PCM the $\hat{A}$ of Eq.~\eqref{eq:interpot} is expressed in terms of operators $\hat{S}$ and $\hat{D}$ which act on surface functions to generate the single- and double-layer potentials, respectively~\cite{mennucci1997evaluation, herbert2021dielectric}. The $\hat{S}$ can be expressed as
\begin{equation}
\label{eq:singleLayer}
\hat{S}\sigma(s)=\int^{}_{\Gamma}ds'\frac{\sigma(s')}{||s'-s||},
\end{equation}
here $s$ denotes a point on the solute cavity surface $\Gamma$. The $\hat{D}$ is defined as
\begin{equation}
\label{eq:doubleLayer}
\hat{D}\sigma(s)=\int^{}_{\Gamma}ds'\sigma(s')\frac{\partial}{\partial n_{s'}}\left( \frac{1}{||s'-s||} \right)
\end{equation}
where $n_{s}$ is the outward-pointing unit vector normal to the cavity surface at the point $s$. Using $\hat{S}$ and $\hat{D}$ the continuum electrostatics problem can be expressed through the integral equation on the surface of the cavity as
\begin{equation}
\label{eq:IEF}
\left[ \left( \frac{2\pi}{f_{\varepsilon}} \right)\hat{I}-\hat{D}\right]\hat{S}\sigma(s) = (-2\pi\hat{I}+\hat{D})\varphi^{\rho}(s),
\end{equation}
here $\hat{I}$ is the identity operator, $\varphi^{\rho}(s)$ is a molecular electrostatic potential evaluated at the cavity surface, and $f_{\epsilon}$ is the permittivity-dependent prefactor.~\cite{herbert2021dielectric} As can be seen from Eq.~\eqref{eq:solventH} and Eq.~\eqref{eq:Gmin} the IEF-PCM is implemented as the expansion of the gas phase Hamiltonian, $\hat{H}^{0}$, by the solute-solvent interaction potential, $\hat{V}_{int}$, which means that to carry out IEF-PCM one needs to first obtain the  $\hat{H}^{0}$ for the method of choice. Hence, in the next section we show that the $\hat{H}^{0}$ can be obtained based on quantum computing simulations, while the expansion by $\hat{V}_{int}$ can be done as classical post-processing step.

While IEF-PCM efficiently accounts for electrostatic interaction between solute and solvent, to achieve the most accurate description of solute-solvent interactions one needs to account for nonelectrostatic contributions as well. The nonelectrostatic interactions between the solute and solvent, including cavitation, Pauli repulsion, dispersion, and hydrogen-bonding can only be efficiently obtained by inclusion of explicit solvent molecules~\cite{herbert2021dielectric}. 

The SMx models~\cite{cramer2008universal} and in particular SMD~\cite{marenich2009universal} are viewed as the most accurate models of implicit solvent allowing for efficient treatment of nonelectrostatic interactions between the solute and solvent~\cite{herbert2021dielectric}. Nonetheless, the focus of the present paper was to demonstrate the application of a continuum solvent model in SQD simulations, where we chose to first explore the use of IEF-PCM which also affords a good representation of solvation effects. Further improvement of solvent models in SQD is the subject of future studies using both implicit and explicit models of solvation.

\bigskip
\tocless\subsection{Sample-based quantum diagonalization with IEF-PCM}

The SQD IEF-PCM simulations start in a similar manner to the standard SQD~\cite{kanno2023quantum,nakagawa2023adapt,robledo2024chemistry} method, where the quantum circuit $| \Phi_{\mathrm{qc}} \rangle$ is executed to sample a set of computational basis states $\chi = \{ \bts{x}_1 \dots \bts{x}_d \}$ from the probability distribution $p(\bts{x}) = | \langle \bts{x} | \Phi_{\mathrm{qc}} \rangle |^2$ of the molecular system in the gas phase. The standard Jordan-Wigner (JW) mapping is used to map fermions to qubits~\cite{ortiz2002simulating,somma2002simulating,somma2005quantum}. Computational basis states are sampled from the truncated version of the local unitary cluster Jastrow (LUCJ) ansatz~\cite{motta2023bridging} expressed as
\begin{equation}
\label{eq:lucj}
| \Phi_{\mathrm{qc}} \rangle = e^{-\hat{K}_2} e^{\hat{K}_1} e^{i\hat{J}_{1}} e^{-\hat{K}_{1}} | {\bf{x}}_{\mathrm{RHF}} \rangle \;,
\end{equation}
where ${\hat{K}}_1$ and ${\hat{K}}_2$ are one-body operators, ${\hat{J}}_1$ is density-density operator, and $| {\bf{x}}_{\mathrm{RHF}} \rangle$ is the restricted closed-shell Hartree-Fock (RHF) state.
The parametrization of the LUCJ ansatz is derived from a classical gas-phase restricted closed-shell CCSD calculations, as was done in the previous SQD studies~\cite{robledo2024chemistry, kaliakin2024accurate}. Contemporary quantum computers unavoidably introduce noise during the execution of an ansatz, which in turn results in noise-corrupted samples with broken particle-number and spin-$z$ symmetries. The percentage of noise-corrupted samples depends on the fidelity of individual devices and the error mitigation techniques applied. 

To restore the particle-number and spin-$z$ symmetries of these noise-corrupted samples SQD employs an iterative self-consistent configuration recovery (S-CORE) procedure~\cite{robledo2024chemistry}.
The S-CORE utilizes: 1) a fixed set of computational basis states $\tilde{\chi}$ sampled from a quantum computer; 2) an approximation to the ground-state occupation number distribution $n_{p\sigma} = \langle \Psi | \crt{p\sigma} \dst{p\sigma} | \Psi \rangle$. S-CORE randomly flips the entries of the computational basis states in $\tilde{\chi}$ using the distance from the current value of the bit and $n_{p\sigma}$. This procedure is carried out until the particle number and spin-$z$ match the target values, which produces the $\tilde{\chi}_R$. $K$ subsets (batches) are pulled from $\tilde{\chi}_R$, which are denoted as $\tilde{\chi}_b$ where $b = 1 \dots K$. Each batch yields a subspace $S^{(b)}$ of dimension $d$~\cite{robledo2024chemistry}. Construction of the subspaces $S^{(b)}$ involves extension of the set of configurations $\tilde{\chi}_b$ to ensure the closure under spin inversion symmetry~\cite{robledo2024chemistry}, which results in larger values of $d$ than $|\tilde{\chi}_b|$. For these subspaces the Hamiltonian is projected as
\begin{equation}
\label{eq6}
\hat{H}_{S^{(b)}}=\hat{P}_{S^{(b)}}\hat{H}\hat{P}_{S^{(b)}}
\;,
\end{equation}
where the projector $\hat{P}_{S^{(b)}}$ is
\begin{equation}
\label{eq7}
\hat{P}_{S^{(b)}} =\sum_{ {\bf{x}} \in S^{(b)}} | {\bf{x}} \rangle \langle {\bf{x}} |
\;.
\end{equation}

Next step of the algorithm is the key part differentiating the SQD IEF-PCM from the standard (gas-phase) SQD. In standard SQD the $\hat{H}_{S^{(b)}}$ is directly employed to perform Davidson diagonalization producing the ground-state wavefunctions, $|\psi^{(b)}\rangle$, and energies, $E^{(b)}$, of the batches. In SQD IEF-PCM after $\hat{H}_{S^{(b)}}$ is formed we use Eq.~\eqref{eq:solventH} (where $\hat{H}_{S^{(b)}} \equiv  \hat{H}^{0}$) to introduce the $\hat{V}_{int}$ in computations of $|\psi^{(b)}\rangle$ and $E^{(b)}$. Importantly, we utilize Eq.~\eqref{eq:Gmin} to solve the SCRF problem. We use the lowest energy across the batches, $\min_b E^{(b)}$, as the best approximation to the ground-state energy, while the $G_{solv}^{(b)}$ of the lowest energy batch is taken as the best approximation to $G_{solv}$. The wavefunctions $|\psi^{(b)} \rangle$ are then employed to update the occupation number distribution,
\begin{equation}
\label{eq8}
n_{p\sigma}=\frac{1}{K} \sum_{1\leq b\leq K } \langle \psi^{(b)} | \crt{p\sigma} \dst{p\sigma} | \psi^{(b)} \rangle
\;,
\end{equation}
where $n_{p\sigma}$ is used as an input in the next S-CORE iteration. To start the S-CORE loop one needs the initial approximation of $n_{p\sigma}$, which in the first iteration of S-CORE is formed from the measurement outcomes with the correct particle number~\cite{huggins2021efficient}.

\tocless\subsection{Computational details}

\paragraph*{\textbf{Geometry optimization, active space selection, and classical benchmark.}} To generate the geometries of methanol, methylamine, ethanol, and water molecules in aqueous solution we perform a geometry optimization at the RHF/cc-pVDZ IEF-PCM level of theory in the PySCF software package~\cite{sun2020recent,sun2018pyscf,sun2015libcint}. The geometry optimization is performed with translation-rotation-internal coordinate (TRIC) system~\cite{wang2016geometry} as implemented in PySCF. The main goal of the present paper is to demonstrate that for a given geometry SQD/cc-pVDZ IEF-PCM simulations can be as accurate as CASCI/cc-pVDZ IEF-PCM simulations, while maintaining reasonable agreement in the predicted $G_{solv}$ compared to the MNSol database~\cite{marenich2020minnesota}. Hence, we opted out from the usage of higher level of theory for the geometry optimizations in favor of computational efficiency. To assess the accuracy of SQD/cc-pVDZ IEF-PCM calculations, we perform CASCI/cc-pVDZ IEF-PCM simulations as implemented in PySCF. The active spaces for methanol, methylamine, and ethanol are constructed using the atomic valence active space (AVAS) method~\cite{Sayfutyarova2017} (as implemented in PySCF) where we select active-space orbitals that overlap with the atomic orbitals (AOs) as listed in column 3 of Table~\ref{table:AS}. The resulting molecular orbitals (MOs) are listed in column 4 of Table~\ref{table:AS}. In the case of the water molecule we simulate all of the orbitals within the cc-pVDZ basis set, excluding only the 1s orbital of the oxygen atom (core MO of oxygen).

\begin{table*}[ht]
\begin{tabular}{llll}
\hline\hline
Species           & Active Space (AS) & Atomic Orbitals (AOs)      & Molecular Orbitals (MOs) \\
\hline
H$_{2}$O          & (8e,23o)          & 1s is excluded                   & core MO of oxygen is excluded \\
CH$_{3}$OH        & (14e,12o)         & C[2s,2p], O[2s,2p], H[1s]   & $\sigma$(C-H, O-H); \\
&&&$\sigma^*$(C-H, O-H); 2 lp(O) \\
C$_{2}$H$_{5}$OH  & (20e,18o)         & C[2s,2p], O[2s,2p], H[1s]   & $\sigma$(C-C, C-H, O-H); \\
&&&$\sigma^*$(C-C, C-H, O-H); 2 lp(O) \\
CH$_{3}$NH$_{2}$  & (14e,13o)         & C[2s,2p], N[2s,2p], H[1s]   & $\sigma$(C-H, N-H); \\
&&&$\sigma^*$(C-H, N-H); 1 lp(N) \\
\hline\hline
\end{tabular}
\caption{Active spaces used in the present study. Active spaces (AS) are described in terms of atomic orbitals (AOs) used in the AVAS procedure and the resulting molecular orbitals (MOs) of each system. Here $\sigma$ denotes bonding MOs, $\sigma^*$ represents antibonding MOs, and $lp$ corresponds to lone pairs.}
\label{table:AS}
\end{table*}

\begin{figure*}
    \centering
    \includegraphics[width=0.8\textwidth]{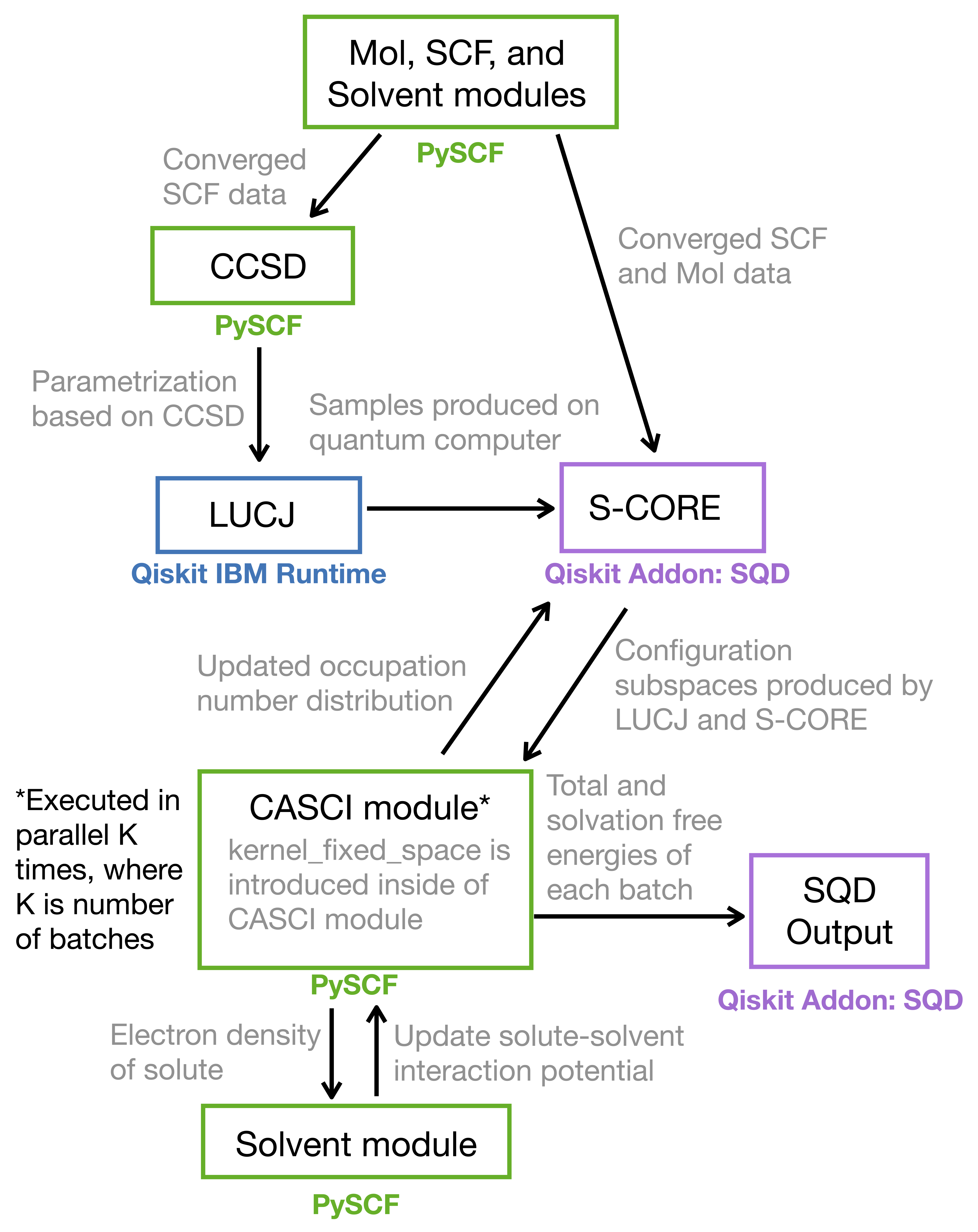}
    \caption{Workflow of the SQD IEF-PCM method. The blue box signifies the part of the workflow that is executed on the quantum computer with Qiskit IBM Runtime. Purple and green boxes indicate the parts of the workflow that are performed with Qiskit Addon: SQD and PySCF subroutines, respectively.}
\label{fig:workflow}
\end{figure*}

\begin{figure*}[t!]
\centering
\includegraphics[width=1.0\textwidth]{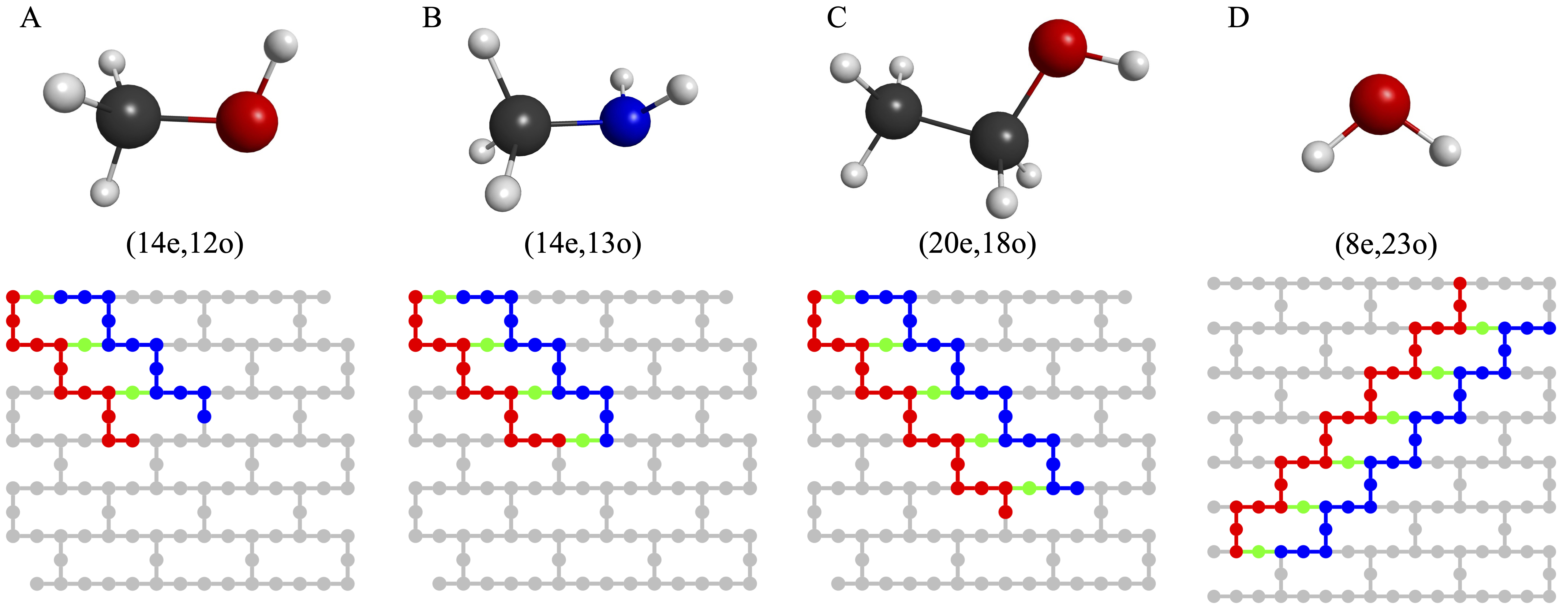}
\caption{Qubit layouts of LUCJ circuits. (A) (14e,12o) simulations of methanol using 27 qubits of \device{cleveland}. (B) (14e,13o) simulations of methylamine using 30 qubits of \device{cleveland}. (C) (20e,18o) simulations of ethanol using 41 qubits of \device{kyiv}. (D) (8e,23o) simulations of water using 52 qubits of \device{marrakesh}. The layouts of quantum devices are shown in gray. Qubits used to encode occupation numbers of spin-up/down electrons are marked in red/blue, while ancilla qubits denoted in green. The structures above the qubit layouts represent the corresponding molecules. Carbon, oxygen, nitrogen, and hydrogen atoms are marked in black, red, blue, and grey, respectively.}
\label{fig:layout}
\end{figure*}

\paragraph*{\textbf{SQD IEF-PCM code implementation.}} The SQD IEF-PCM method is enabled through modification of the Qiskit addon: SQD~\cite{sqd_addon}  and PySCF~\cite{pyscf_github} codes. PySCF has a dedicated "$CASCI$" module which incorporates the classical full configuration interaction (FCI) and selected configuration interaction (SCI) solvers. For these classical solvers the $"CASCI"$ module of PySCF has a well-established integration with the $"solvent"$ module of PySCF~\cite{sun2020recent}. The standard (gas phase) SQD code bypasses the $"CASCI"$ module of PySCF and instead directly accesses the "$kernel\_fixed\_space$" data structure of PySCF to perform Davidson diagonalization in the subspace produced by LUCJ and S-CORE. The "$kernel\_fixed\_space$" data structure of standard PySCF is lacking the interface with the $"solvent"$ module.

To enable implicit solvent functionality, we introduce two modifications in the "$solve\_fermion$" function of SQD: A) "$solve\_fermion$" receives an additional input argument encapsulating the solvent model and all of the associated data; B) "$solve\_fermion$" performs the call to the $"CASCI"$ module of PySCF instead of a direct call to the "$kernel\_fixed\_space$" data structure. On the PySCF side we modify the $"CASCI"$ module as follows: A) "$kernel\_fixed\_space$" is incorporated inside of the $"CASCI"$ module of PySCF; B) the $"CASCI"$ module of PySCF receives an additional input argument containing the subspace produced by the LUCJ and S-CORE procedures. Finally, we also modify both the "$solvent$" and $"CASCI"$ modules of PySCF to include $G_{solv}^{(b)}$ as one of the output arguments of the $"CASCI"$ module. Later modification allows for the passing of $G_{solv}^{(b)}$ as one of the "$solve\_fermion$" return arguments in SQD. 

Our code modifications enable access to not only the IEF-PCM implicit solvent model, but to all of the solvent models encapsulated in the "$solvent$"  module of PySCF. However, further tests of other solvent models is outside of the scope of the present paper, where our main goal was to show that we can perform implicit solvent simulations using the SQD method with one of the popular implicit solvent models. The workflow of the SQD IEF-PCM method is summarized in Fig.~\ref{fig:workflow}.

\begin{figure*}
    \centering
    \includegraphics[width=0.5\textwidth]{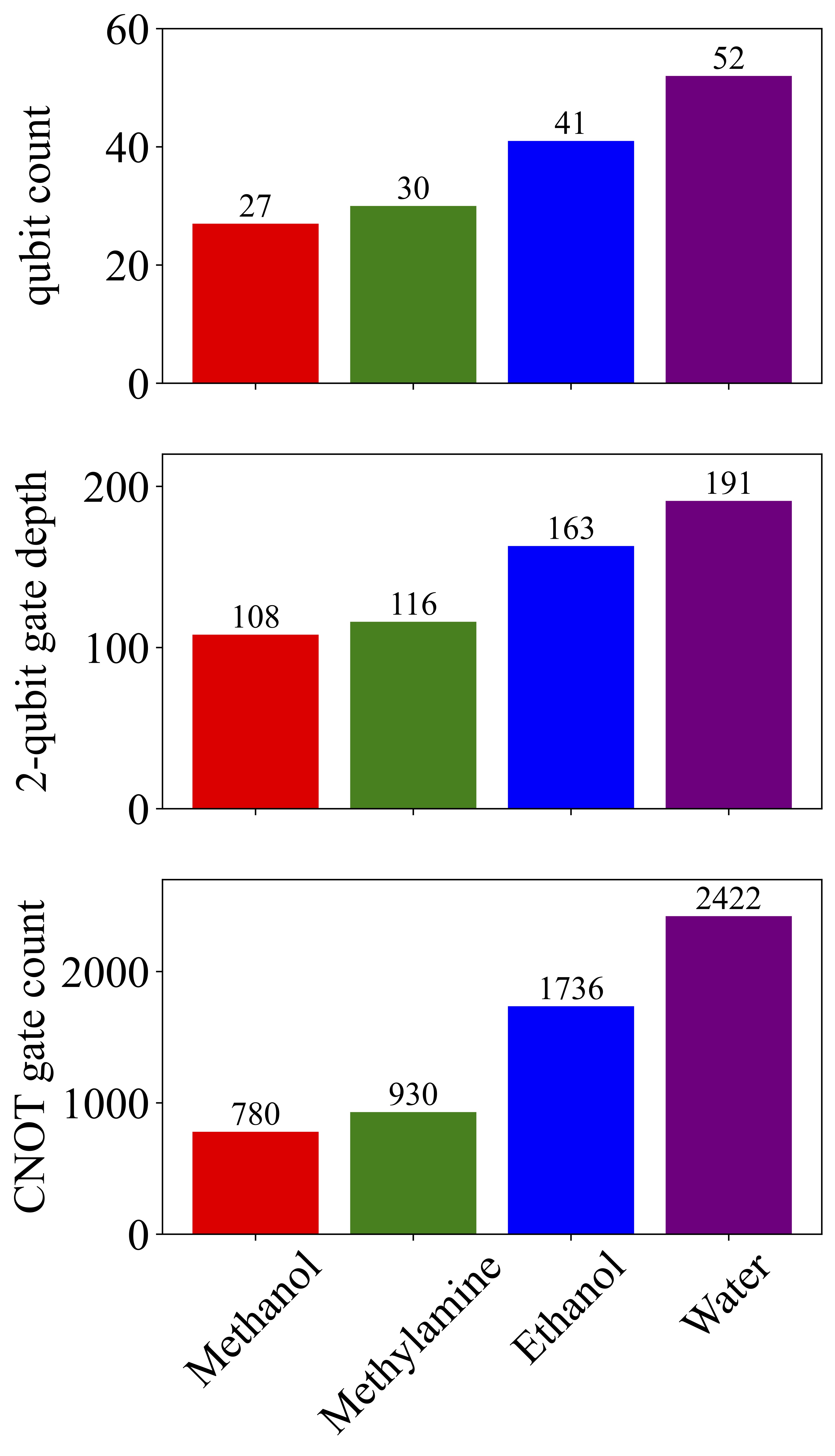}
    \caption{Scaling of qubits and quantum gate operations. Number of qubits, 2-qubit gate depth, and CNOT gate count for LUCJ circuits of methanol (14e,12o), methylamine (14e,13o), ethanol (20e,18o), and water (8e,23o) calculations, represented by red, blue, green, and purple columns, respectively.}
\end{figure*}

\paragraph*{\textbf{LUCJ and SQD simulations.}} The LUCJ quantum circuits are generated using the ffsim library~\cite{ffsim2024} interfaced with Qiskit~\cite{aleksandrowicz2019qiskit,javadi2024quantum}. The quantum circuits are executed on the \device{cleveland}, \device{kyiv}, and \device{marrakesh} quantum computers, using the qubit layouts represented in Fig.~\ref{fig:layout}A, \ref{fig:layout}B, \ref{fig:layout}C, and \ref{fig:layout}D. The mitigation of quantum errors is done through gate twirling (but not measurement twirling) over random 2-qubit Clifford gates~\cite{wallman2016noise} and dynamical decoupling~\cite{viola1998dynamical,kofman2001universal,biercuk2009optimized,niu2022effects} as available via the SamplerV2 primitive in the Qiskit's runtime library. The number of qubits, 2-qubit gate depth, and number of CNOT gates in the LUCJ circuits are shown in Fig.~\ref{fig:layout}. In SQD/cc-pVDZ IEF-PCM calculations we utilize 3 iterations of S-CORE and 10 subsets (batches). We use parallelization across 10 CPUs with Ray~\cite{moritz2017ray} where the eigenstate solver within each of the 10 batches is using 1 CPU. The details regarding the number of samples and configurations used in SQD IEF-PCM calculations are listed in Table~\ref{table:sqd}.

\begin{table*}[t!]
\begin{tabular}{lllll}
\hline\hline
system  & AS & $|\tilde{\chi}_b|$ $[10^3]$ & $d$ $[10^5]$  & $D_{\mathrm{AS}}$ $[10^5]$   \\
\hline
CH$_{3}$OH        & (14e,12o)    & 0.2      & 0.745        & 6.273  \\
CH$_{3}$OH        & (14e,12o)    & 0.4      & 1.832        & 6.273  \\
CH$_{3}$OH        & (14e,12o)    & 0.6      & 2.652        & 6.273  \\
CH$_{3}$OH        & (14e,12o)    & 0.8      & 3.481        & 6.273  \\
CH$_{3}$OH        & (14e,12o)    & 1.0      & 4.020        & 6.273  \\
CH$_{3}$NH$_{2}$  & (14e,13o)    & 0.5      & 4.134        & 29.447  \\
CH$_{3}$NH$_{2}$  & (14e,13o)    & 1.0      & 9.960        & 29.447  \\
CH$_{3}$NH$_{2}$  & (14e,13o)    & 1.5      & 14.520       & 29.447  \\
CH$_{3}$NH$_{2}$  & (14e,13o)    & 2.0      & 18.333       & 29.447  \\
CH$_{3}$NH$_{2}$  & (14e,13o)    & 2.5      & 21.199       & 29.447  \\
C$_{2}$H$_{5}$OH  & (20e,18o)    & 10.0     & 1960.000     & 19147.626  \\
C$_{2}$H$_{5}$OH  & (20e,18o)    & 12.0     & 2549.451     & 19147.626  \\
C$_{2}$H$_{5}$OH  & (20e,18o)    & 14.0     & 3162.351     & 19147.626  \\
C$_{2}$H$_{5}$OH  & (20e,18o)    & 16.0     & 3796.263     & 19147.626  \\
C$_{2}$H$_{5}$OH  & (20e,18o)    & 18.0     & 4445.351     & 19147.626  \\
H$_{2}$O          & (8e,23o)     & 6.0      & 155.078      & 784.110   \\
H$_{2}$O          & (8e,23o)     & 8.0      & 215.018      & 784.110   \\
H$_{2}$O          & (8e,23o)     & 10.0     & 272.380      & 784.110   \\
H$_{2}$O          & (8e,23o)     & 12.0     & 309.247      & 784.110   \\
H$_{2}$O          & (8e,23o)     & 14.0     & 352.005      & 784.110   \\
\hline\hline                       
\end{tabular}
\caption{Details of SQD/cc-pVDZ IEF-PCM calculations. $AS$ and $D_{\mathrm{AS}}$ are abbreviations for active space and Hilbert-space dimension, respectively. The value of $d$ in column 4 corresponds to the subset (batch) with the lowest energy across batches at the last iteration of S-CORE.
}
\label{table:sqd}
\end{table*}


\tocless\section{Results and Discussion}

\begin{figure*}[!ht]
    \centering
    \includegraphics[width=1\textwidth]{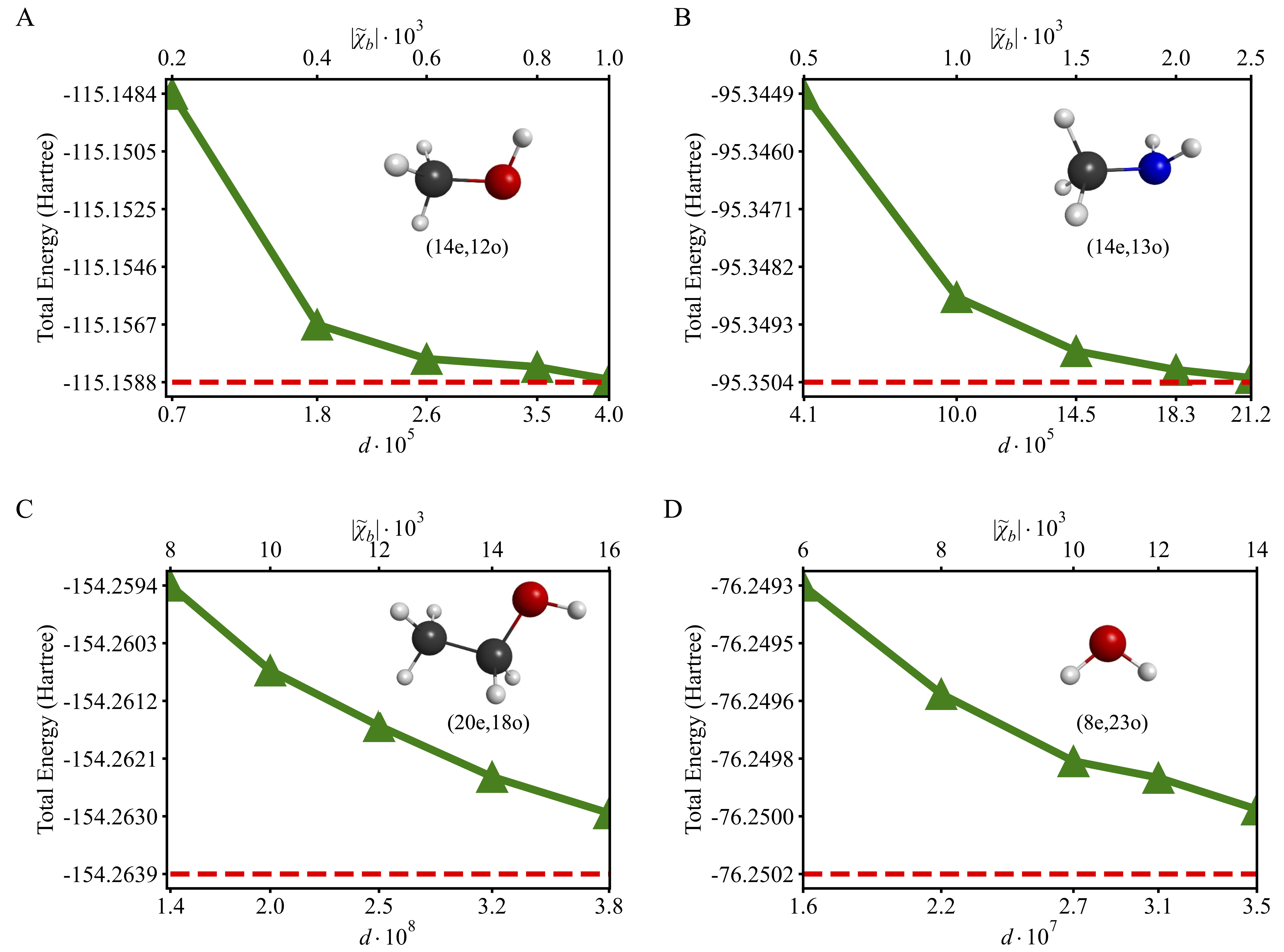}
    \caption{Total energy of solvated molecules as a function of $d \cdot 10^x$, where $x$ varies for each simulation: (A) (14e,12o) methanol with $x=5$, (B) (14e,13o) methylamine with $x=5$, (C) (20e,18o) ethanol with $x=8$, and (D) (8e,23o) water with $x=7$. The secondary x-axis demonstrates the value of $|\widetilde{\chi}_b| \cdot 10^3$ producing the given value of $d \cdot 10^x$. The solid green line with triangular markers shows SQD/cc-pVDZ IEF-PCM results. The horizontal dashed red line indicates the reference total energy from CASCI/cc-pVDZ IEF-PCM calculations.}
\end{figure*}

Figure 4 illustrates the total energy of four simulations of solvated molecules, namely (14e,12o) methanol (CH\(_3\)OH), (14e,13o) methylamine (CH\(_3\)NH\(_2\)), (20e,18o) ethanol (C\(_2\)H\(_5\)OH), and (8e,23o) water (H\(_2\)O), as a function of $d$. The results, obtained using the SQD/cc-pVDZ IEF-PCM approach, are compared with reference CASCI/cc-pVDZ IEF-PCM energies. For all systems, as the sample size increases, the total energy computed with SQD IEF-PCM converges systematically toward the reference energy.

Methanol, shown in Fig. 4.A, with an active space of (14e,12o) and a Hilbert space dimension of \(6.27 \times 10^5\), achieves rapid convergence due to its smaller state space. At the lowest sample size of \(|\tilde{\chi}_b|=0.2 \cdot 10^3\), the SQD IEF-PCM energy deviates by 6.51 kcal/mol from CASCI IEF-PCM. This reduces to just 0.06 kcal/mol at \(|\tilde{\chi}_b|=1.0 \cdot 10^3\) corresponding to \(ca. \ 66\%\) of the Hilbert space.

For methylamine, shown in Fig. 4.B, with an active space of (14e,13o) and a Hilbert space dimension of \(2.94 \times 10^6\), the energy difference decreases from 3.45 kcal/mol at the lowest sample size of \(|\tilde{\chi}_b|=0.5 \cdot 10^3\) to 0.05 kcal/mol at the highest sample size of \(|\tilde{\chi}_b|=2.5 \cdot 10^3\). This corresponds to increase from \(ca. \ 14\%\) to $ca.$ 70\% of the Hilbert space, demonstrating convergence with an increase of the sample size.

Ethanol, shown in Fig. 4.C, with an active space of (20e,18o) and a much larger Hilbert space dimension of \(1.91 \times 10^9\), requires significantly larger sample sizes for convergence. At the lowest sample size of \(|\tilde{\chi}_b|=10.0 \cdot 10^3\), the SQD IEF-PCM energy is 1.97 kcal/mol higher than the CASCI IEF-PCM reference. At the largest sample size of \(|\tilde{\chi}_b|=18.0 \cdot 10^3\), this discrepancy reduces to 0.34 kcal/mol, reflecting improved resolution of its electronic structure. However, even the largest sample covers only \(ca. \ 23\%\) of the Hilbert space, indicating that for this system SQD IEF-PCM efficiently samples the most dominant configurations, while producing results within chemical accuracy.

For water, shown in Fig. 4.D, with an active space of (8e,23o), the Hilbert space dimension is \(7.84 \times 10^7\). At a sample size of \(|\tilde{\chi}_b|=6.0 \cdot 10^3\), the energy deviates by 0.59 kcal/mol from the CASCI IEF-PCM reference, while at \(|\tilde{\chi}_b|=14.0 \cdot 10^3\), the energy difference decreases to 0.13 kcal/mol. This highlights the growing coverage of the Hilbert space, from approximately 13\% at the lowest sample size to nearly 45\% at the highest.

Overall, the SQD IEF-PCM method achieves excellent agreement with CASCI IEF-PCM across all systems, with the largest discrepancies at low sample sizes. Increasing the sample size systematically reduces these deviations, exceeding chemical accuracy at higher sampling rates, even for more complex systems like ethanol. This demonstrates the potential of the SQD IEF-PCM approach to deliver chemically accurate energies for solvated molecules while efficiently sampling the relevant portions of Hilbert space.

\begin{table*}[ht]
    \centering
    \begin{tabular}{lllll}
    \hline\hline
      System   & $SQD_{lns}$ IEF-PCM & $SQD_{hns}$ IEF-PCM & CASCI IEF-PCM & MNSol\\ \hline
        CH$_3$OH (14e,12o) & -4.67 & -4.51 & -4.50 & -5.11\\ 
        CH$_3$NH$_2$ (14e,13o) & -4.01 & -3.99  & -3.99 & -4.56\\ 
        C$_2$H$_5$OH (20e,18o) & -4.46 &-4.42  &-4.42  &-5.01 \\ 
        H$_2$O (8e,23o) & -6.18 & -6.15 & -6.15& -6.31\\
    \hline\hline
    \end{tabular}
    \caption{Solvation free energies. $G_{solv}$ for studied systems calculated using SQD/cc-pVDZ IEF-PCM and CASCI/cc-pVDZ IEF-PCM in comparison with the MNSol database~\cite{marenich2020minnesota}. For SQD/cc-pVDZ IEF-PCM calculations we also demonstrate how large the fluctuations of the predicted $G_{solv}$ is in the case of the lowest and highest number of samples. Here $lns$ denotes the lowest number of samples and $hns$ denotes the highest number of samples.}
    \label{tab:my_label}
\end{table*}

Table 3 presents the solvation free energies (\(G_{solv}\)) for the solvated molecules studied, namely methanol (CH\(_3\)OH), methylamine (CH\(_3\)NH\(_2\)), ethanol (C\(_2\)H\(_5\)OH), and water (H\(_2\)O), calculated using the SQD IEF-PCM and CASCI IEF-PCM methods, along with comparisons to the MNSol database~\cite{marenich2020minnesota}. The results show that for both the lowest and highest sample cases, the SQD IEF-PCM calculations reproduce the $G_{solv}$ of CASCI IEF-PCM almost exactly (within 0.04 kcal/mol) with the exception of methanol (14e,12o) SQD IEF-PCM simulation utilizing 200 samples per batch where the difference in $G_{solv}$ between two methods is 0.16 kcal/mol. Moreover, the deviation between the SQD IEF-PCM, CASCI IEF-PCM and MNSol solvation free energies is consistently below 1 kcal/mol for all systems, further confirming the accuracy of the SQD approach coupled with IEF-PCM in predicting solvation free energies.

In terms of the solvation free energy values, the SQD IEF-PCM method shows minor fluctuations between the lowest and highest sample sizes, but these deviations are small and do not significantly impact the overall accuracy. Hence, the SQD IEF-PCM method proves to be a reliable and efficient tool for the calculations of solvation free energies, closely matching the results from higher-level methods like CASCI IEF-PCM while maintaining computational efficiency.


\tocless\section{Conclusion}
In the present work we implemented SQD/cc-pVDZ IEF-PCM simulations using the modified PySCF and Qiskit Addon: SQD codes. We deployed our SQD/cc-pVDZ IEF-PCM simulations on real quantum hardware utilizing 27, 30,
41, and 52 qubits for (14e,12o) methanol, (14e,13o) methylamine, (20e,18o) ethanol, and (8e,23o) water models in aqueous solution. We demonstrate that with sufficient sampling the SQD/cc-pVDZ IEF-PCM total energies agree with CASCI/cc-pVDZ IEF-PCM results within 0.06, 0.05, 0.35, and 0.13 kcal/mol for (14e,12o) methanol, (14e,13o) methylamine, (20e,18o) ethanol, and (8e,23o) water simulations, respectively. This work is the first demonstration of implicit solvent simulations with SQD, which is a promising new avenue for quantum simulations of biologically-relevant chemical reactions and drug discovery.


\begin{acknowledgement}
The authors gratefully acknowledge financial support from the National Science Foundation (NSF) through CSSI Frameworks Grant OAC-2209717 and from the National Institutes of Health (Grant Numbers GM130641). The authors also thank Javier Robledo Moreno, Mario Motta, Caleb Johnson, Abdullah Ash Saki, Iskandar Sitdikov, and Antonio Mezzacapo for their guidance on the original SQD method. The close support of IBM on the original SQD method was critical for our ability to create the SQD IEF-PCM methodology.
\end{acknowledgement}

\newpage
\bibliography{SQD-PCM}

\end{document}